# Giant magnetic quantum oscillations in the thermal conductivity of TaAs: Indications of chiral zero sound

[1,2] Junsen Xiang, [1,3] Sile Hu, [4,1] Zhida Song, [1,3] Meng Lv, [1,3] Jiahao Zhang, [1,3] Lingxiao Zhao, [2] Wei Li, [2] Ziyu Chen, [1] Shuai Zhang, [1,3] Jiantao Wang, [1,3] Yi-feng Yang, [5] Xi Dai, [1,6] Frank Steglich, [1,3,7] Genfu Chen & [1,3,7] Peijie Sun

*1 Beijing National Laboratory for Condensed Matter Physics, Institute of Physics, Chinese Academy of Sciences, Beijing 100190, China*
*2 Department of Physics, Key Laboratory of Micro-Nano Measurement-Manipulation and Physics, Beihang University, Beijing 100191, China*
*3 University of Chinese Academy of Sciences, Beijing 100049, China*
*4 Department of Physics, Princeton University, Princeton, NJ 08544, USA*
*5 Department of Physics, Hong Kong University of Science of Technology, Clear Water Bay, Kowloon, Hong Kong*
*6 Max Planck Institute for Chemical Physics of Solids, 01187 Dresden, Germany*
*7 Songshan Lake Materials Laboratory, Dongguan, Guangdong 523808, China*

**Abstract:**
**Charge transport of topological semimetals has been in the focus of intensive investigations because of their non-trivial band topology. Heat transport of these materials, on the other hand, is largely unexplored and remains elusive. Here we report on an observation of unprecedented, giant magnetic quantum oscillations of thermal conductivity in the prototypical Weyl semimetal TaAs. The oscillations are antiphase with the quantum oscillating electronic density of states of a Weyl pocket, and their amplitudes amount to two orders of magnitude of the estimation based on the Wiedemann-Franz law. Our analyses show that all the conventional heat-transport mechanisms through diffusions of propagating electrons, phonons and electron-hole bipolar excitations, are far inadequate to account for these phenomena. Taking further experimental facts that the parallel field configuration favors much higher magneto-thermal conductivity, we propose that the newly proposed chiral zero sound provides a reasonable explanation to these exotic phenomena. More work focusing on other topological semimetals along the same line is badly called for.**

Charge and heat conductions define two entangled fundamental transport properties of a conducting solid. Controllable manipulations of both are essential for various functionalities of pertinent materials. For the recently discovered topological semimetals, electrical transport has revealed distinct topologically nontrivial properties, like the chiral anomaly in the longitudinal magneto-resistance (*MR*) [1-4]. On the other hand, thermal transport, in particular thermal conductivity, has remained largely unexplored. This is mainly due to the technical difficulties of precise thermal management at low temperatures. In addition, such work has been so far considered uninteresting because the thermal conductivity may be estimated from the electrical conductivity through the Wiedemann-Franz (WF) law.

Indeed, for most electrical conductors, the WF law can provide a straightforward and reliable approach to the electronic contribution $\kappa_e$ of the thermal conductivity based on the electrical conductivity $\sigma$,

$$\kappa_e / T = \sigma L_0, \quad (1)$$

with the Sommerfeld value of Lorenz number $L_0 \equiv \pi^2/3 \, (k_B/2)^2 = 2.44 \times 10^{-8}$ W·Ω·K$^{-2}$. The WF law holds in the case of dominant elastic scattering, e.g., from static defects as realized at sufficiently low temperatures [5]. Within the last few years, it was also found that the charge and heat conductions of the relativistic Dirac/Weyl fermions may deviate from the WF law [6-11]. For example, it was shown very recently that hydrodynamic transport by relativistic Weyl fermions in a Weyl semimetal WP$_2$ can lead to $L/L_0 < 1$ in the temperature range $T < 200$ K [10]. On the other hand, for the electron-hole plasma at the charge-neutrality point in graphene, considered to form a strongly coupled Dirac fluid, hydrodynamic transport was found between $T = 50$ and 80 K to result in a much enhanced Lorenz number $L \approx 22 \, L_0$ [11].

Large magnetic quantum oscillations (MQOs) in thermal transport properties, like Seebeck and Nernst effect, have recently been detected and found to complement the conventional techniques like the Shubnikov-de Haas (SdH) effect of $\sigma(B)$ [12-15]. In contrast, the thermal conductivity $\kappa(B)$ has only rarely been used in this respect. An estimate of the MQOs in $\kappa_{e,WF}(B)$, the electronic thermal conductivity derived from $\sigma(B)$ utilizing the WF ratio $L_0$, are generally too small to be detected. The very few examples of $\kappa(B)$ MQOs hitherto reported include a bismuth single crystal [16] and GaAs/AlGaAs heterostructures containing a two-dimensional electron system (2DES) [17]. Here, the amplitude of the MQOs compared to the respective zero-field thermal conductivity $\kappa_0$ is less than 8 %. However, for the Weyl semimetal NbP [18], where $\kappa(B)$ MQOs are also observed, it was found to amount to about 30% of $\kappa_0$. A conventional electronic contribution in accord with the WF law, a modulated lattice thermal conductance through electron-phonon (e-p) interaction in a quantum oscillating 2DES, or a large electronic bipolar-diffusion conduction, respectively, have been discussed as possible origins of this remarkable observation.

In this work, we report the experimental observation of unprecedentedly large MQOs in the magneto-thermal conductivity $\kappa(B)$ of TaAs, a prototypical, recently identified Weyl semimetal [19-22]. The $\kappa(B)$ MQOs exhibit a frequency ($F \approx 7$ T for $B \parallel c$) which is characteristic for a Weyl fermion pocket that has already been confirmed in the bulk state of TaAs [19,22]. The amplitude of these MQOs turns out to be surprisingly large; it amounts to more than 300% of the value of $\kappa_0$ when the magnetic field is aligned along the thermal current, i.e., $B \parallel dT \parallel c$. Moreover, these oscillations are *antiphase* to those observed in the electrical conductivity $\sigma(B)$. Our detailed analysis indicates that such giant $\kappa(B)$ MQOs are incompatible with any well-established heat conduction mechanism in solids, such as the diffusion contributions by phonons and conduction electrons as well as the contribution by bipolar, electron-hole symmetric excitations near a charge neutrality condition. Here, we show that the main features of the observed $\kappa(B)$ MQOs in TaAs can be ascribed to the recently proposed neutral bosonic excitation, i.e., field-induced chiral zero sound (CZS) associated with the exotic chiral magnetic effect in Weyl semimetals [23].

We have employed two TaAs samples cut from the same single crystal grown by chemical vapor transport [19]. They were rectangular shaped, with the long edge either along the *c* axis (denoted as TAc, dimension 0.3 x 0.8 x 2.2 mm$^3$, cf. Fig. 1a) or along the *a* axis (TAa, 0.4 x 1.0 x 2.1 mm$^3$). Thanks to the large sample size that allows for configuring multiple thermal/electrical contacts and in order to reduce analytical errors, simultaneous measurements of electrical resistivity $\rho$ and Seebeck coefficient $S$ were carried out in addition to $\kappa$. Upon scanning the magnetic field, the temperature difference d$T(B)$ across the sample oscillates significantly, see Fig. 1b. In view of the constant heating power $Q$ at the high-$T$ end of the sample, this signifies pronounced quantum oscillations of $\kappa(B)$, see Fig. 2b. The latter were cross-checked by employing two types of temperature sensors, i.e., a field-calibrated thermocouple and a pair of Cernox thermometers. The results obtained this way are in very good agreement, see Fig. S2 in Supplementary Material (SM) [24]. As shown in Fig. 1b, the signal of d$T(B)$ is nearly symmetric for positive and negative fields, hinting at good thermal stability during the whole measurement process. In order to eliminate the weak field-odd, thermal Hall components, we have taken the mean value of d$T(B)$ measured for positive and negative fields throughout this paper. The value of $\kappa$ measured at $B = 0$ fits well to our previous results [25].

In Fig. 2, we compare the isothermal electrical and thermal conductivities $\sigma_\parallel(B)$ (**a**) and $\kappa_\parallel(B)$ (**b**) of sample TAc. The subscript '$\parallel$' denotes a parallel alignment of the electrical/thermal current and the magnetic field. Note that in a parallel field configuration there is no Hall conductivity, and $\sigma_\parallel(B) = 1/\rho_\parallel(B)$. Here $\sigma_\parallel(B)$ rather than $\rho_\parallel(B)$ is employed in order to facilitate the following discussions. The strong decrease of $\sigma_\parallel(B)$ at low fields (MR > 0) changes to a weak increase (MR < 0) at $B$ > 1 T, characteristic of the chiral anomaly arising from charge pumping between Weyl nodes of different chirality [19]. Significant MQOs of $\sigma_\parallel(B)$ at higher fields $B$ > 2 T reveal several characteristic frequencies, i.e., $F \approx$ 7, 15, and 23 T, that are consistent with current knowledge on this material [19-22], cf. inset of Fig. 2 for the fast Fourier transform (FFT) spectrum. The dominating frequency ($F$ = 7 T) is due to a tiny hole-type Fermi pocket enclosing a Weyl node [19]. Both the magnetization $M(B)$ and the simultaneously measured $S_\parallel(B)$ show marked MQOs as well, with the same dominating frequency as derived from electrical conductivity, see Figs. S3 & S4 [24].

As shown in Fig. 2, the MQOs in $\kappa_\parallel(B)$ are substantially more pronounced, especially at $B$ < 4 T, than those in $\sigma_\parallel(B)$. The largest change in the last MQO period of $\kappa_\parallel(B)$ (Fig. 2b) amounts to d$\kappa$ = 12 W/Km, i.e., 3.4 times the value of $\kappa_\parallel$ ($B$ = 0). The MQOs with $F$ $\approx$ 7 T are practically the only visible oscillating component in $\kappa_\parallel(B)$, cf. $\kappa_\parallel(B)/T$ vs $B^{-1}$ shown in Fig. 3a and the corresponding FFT analysis in Fig. S7 [24], in striking contrast to $\sigma_\parallel(B)$ (Fig. 2a) and $S_\parallel(B)$ (Fig. S3 [24]). The excellent agreement of this frequency with that of the Weyl pocket probed by other techniques [19, 21-22] clearly indicates that the MQOs observed in $\kappa_\parallel(B)$ are a consequence of the Landau-quantized electronic density of states (DOS) of the Weyl fermions in the *bulk* of TaAs. MQOs of the same frequency, with 5-6 times smaller amplitude, have also been observed in the perpendicular field configuration, i.e., $\kappa_\perp(B)$ of sample TAa measured with d$T \parallel a$ and $B \parallel c$, see Fig. S5 [24]. Nevertheless, these oscillations are still quite large, when compared to $\kappa_{e,WF}(B)$ estimated from the WF law, cf. Fig. S6 [24]. The recently observed MQOs in $\kappa_\perp(B)$ of NbP ($\kappa_\parallel(B)$ was not measured) reveal a somewhat smaller but still comparable oscillation amplitude (d$\kappa/\kappa_{0T} \approx$ 30%) at $T$ = 2 K [18]. Note that for this compound the FFT spectrum is much more complicated than for TaAs.

The common description of the thermal conductivity in nonmagnetic conductors considers diffusion contributions from both electrons and phonons, i.e., $\kappa_e$ and $\kappa_{ph}$. For semimetals and semiconductors a bipolar conduction term $\kappa_{bi}$ has to be taken into account, too. This is due to thermally excited electron-hole pairs which diffuse to the cold end where they recombine, releasing their excitation energies [26]. As already mentioned in the introduction, $\kappa_e$ can be readily estimated from $\sigma$ via the WF law (Eq. 1) for a wide variety of conducting materials, under dominating elastic scattering [5]. This is often realized at low enough temperatures with dominating impurity scattering. On the other hand, the lattice contribution $\kappa_{ph}$ can be approached from first-principle calculations [27], as long as the intrinsic phonon-phonon Umklapp scattering is the dominating phonon relaxation process. It may also interact with the electron-diffusion term through *e-p* coupling, as has been argued for the Landau-quantized 2DES [17]. A bipolar contribution $\kappa_{bi}$ will become significant only near a charge neutrality condition, as realized, e.g., in graphene [28].

To identify the potential origin of the giant $\kappa(B)$ MQOs in TaAs, we first examine the possibility of a large electronic contribution, $\kappa_e(B)$, that is well enhanced over its WF counterpart $\kappa_{e-WF}(B)$, implying $L(B) >> L_0$. In order for this scenario to work, $L(B)$ has to exceed $L_0$ by a factor of ~100, cf. Fig. 2**b** (Note the logarithmic *y* axis). A similar conclusion can be drawn from measurements with the perpendicular field configuration (Fig. S6 [24]). This situation is different from the chiral-anomaly induced negative *MR* that appears only in parallel fields. In fact, for nonmagnetic conductors, application of a magnetic field hardly affects the Lorenz number *L* because the charge-relaxation mechanisms involved remain unchanged. In Cu, for example, *L* changes at 4.8 K by only 10% when *B* is increased up to 5 T [29]. Recently, electrical and thermal signatures of a hydrodynamic electron fluid with violated WF law have been identified in the topological semimetals like $WP_2$ [10]. There, however, the Lorenz number *L* was found to be largely reduced from $L_0$, contrary to our observations made in TaAs.

There is compelling evidence against the electron-diffusion scenario described above based on the following two observations. The first one concerns the $T$ dependence of the oscillations as shown in Figs. 3**a** & 3**b**, see also the corresponding FFT spectrum shown in Fig. S7 [24]. While the oscillation amplitude of both the magnetization $M(B)$ and $\sigma(B)$ follows the Lifshitz-Kosevich function for thermally damped Landau quantization [22], i.e., steadily decreases upon warming, that of $\kappa_\parallel(B)/T$ does not, cf. Fig. 3**b**. The latter becomes largest at $T \approx 2.3$ K and diminishes upon further cooling, suggesting a non-electronic, most likely bosonic contribution. The second observation is that of an oscillation phase of the observed $\kappa_\parallel(B)$ MQOs which is opposite to that of $\sigma_\parallel(B)$ or $\kappa_{e,WF}(B)$. For example, at $B = 7.3$ T (dashed line in Fig. 2), where $\sigma_\parallel(B)$ shows a maximum because of the intersection of a Landau level with the Fermi level (causing a maximum in the DOS), $\kappa_\parallel(B)$ reveals a minimum (Fig. 2**b**). This implies that in this very field where TaAs is electrically most conductive it is thermally almost insulating, contrary to what is expected in the case of conduction by electron diffusion.

An alternative explanation for the giant MQOs in $\kappa(B)$ may involve the bipolar-diffusion term $\kappa_{bi}(B)$, as argued in the case of NbP [18]. This effect has been intensively investigated for graphene which fulfills all the requirements for bipolar transport, e.g., zero band gap and charge neutrality [28]. There, under proper electrical gating, bipolar diffusion can enhance $L$ by a factor 2 - 4 at room temperature. This factor diminishes with lowering $T$ due to the freezing out of thermal excitations. Experimentally, the enhancement of $L(T)$ was estimated to be only 35% in the subkelvin region [30]. Related to the bipolar excitations in graphene, as already mentioned in the introduction, a further enhancement of $L$ up to $22L_0$ was recently observed at elevated temperatures, i.e., between 50 and 80 K [11]. There, the charge neutral electron-hole plasma obeying hydrodynamics was argued to be realized. In TaAs, the dominating Weyl node is located 2 meV above the Fermi level which intersects with several other topological trivial and nontrivial pockets. This implies that TaAs does not fulfill the charge neutrality requirement for significant transport by electron-hole excitons. While a dominating bipolar heat transport should rapidly diminish upon increasing magnetic field [26] because the electron-hole excitons are easily destroyed, the giant $\kappa(B)$ MQOs observed in TaAs indeed persist up to 9 T, see Fig. 2**b**. This yields further evidence against the bipolar-diffusion scenario.

Having discarded both the electronic and bipolar contributions to heat transport as potential origins of the giant MQOs observed in $\kappa(B)$, we now turn to the possible influence of the phonon contribution $\kappa_{ph}(B)$ for which *e-p* coupling may cause an oscillating behavior due to the field-induced Landau quantization of the electronic DOS at the Fermi energy. This scenario was suggested to explain the relatively small (2%) MQOs of $\kappa(B)$ reported for GaAs/AlGaAs heterostructures [17]. In this case, the phonon mean-free path is modulated in magnetic field, via *e-p* coupling, by the Landau-quantized 2D-DOS which is confined to the interface. This effect should be much weaker in a 3D system with multiple Fermi pockets like TaAs, where discrete Landau levels are absent. To further explore this possibility, we have performed ultrasound measurements on TaAs. In the configuration with the phonon propagation and magnetic field being parallel along *c* axis, the MQOs of the phonon velocity $v(B)$, while also having a dominating frequency of 7 T, are only of the order of $1/10^4$, see Fig. 4, left axis. This is consistent with our first-principles calculations of the phonon-dispersion relations of TaAs which reveal negligible field dependence, see Fig. S8 [24]. Moreover, up to $B$ = 9 T, the observed MQOs of the ultrasound echo amplitude, which are a measure of phonon mean-free path, are only within 5%, see Fig. 4, right axis. Here, we adopted a transverse phonon of relatively low frequency ($f$ = 19 MHz); similar results have been reported for this material by using longitudinal phonons of much higher frequencies (>300 MHz) [31].

Based on our ultrasound measurement results, employing the kinetic formula for the heat conductivity carried by acoustic phonons, $\kappa_{ph}=1/3 C_{ph} v_{ph} l_{ph}$, the magneto-thermal conductivity $\kappa_{\parallel}(B)$ can be estimated to oscillate within 5% of its zero-field value, assuming $\kappa_{ph}(B)$ dominates $\kappa_{\parallel}(B)$. Here $C_{ph}$ is the phonon contribution the specific heat, $v_{ph}$ the phonon velocity, and $l_{ph}$ the phonon mean-free path. While this may indeed account for the $\kappa(B)$ MQOs observed in bismuth [16], it is much too small to explain our observations on TaAs. Moreover, the smooth background of the ultrasound echo amplitude tends to decrease as a function of $B$ (see dashed blue line in Fig. 4). A similar tendency can also be derived from the ultrasound attenuation at $B$ < 10 T as reported in ref. 31. These are in stark contrast to our observation that the smooth background $\kappa_{bg}(B)$ of $\kappa_{\parallel}(B)$ increases with field (Fig. 5, to be discussed below), although both phonon propagation and heat transport were measured with the same parallel field along *c* axis. To conclude, though the propagating acoustic phonons can indeed be affected by the Landau-quantized electronic DOS and give rise to MQOs of both sound velocity and echo amplitude, they are unapt to account for the giant, experimentally observed MQOs of the thermal conductivity of TaAs.

Recently, a distinct magneto-thermal conductivity arising from the Fermi arcs at the *surfaces* of Weyl semimetals, mediated by Weyl node, has been proposed [32]. However, this also cannot explain our observations because the giant $\kappa(B)$ MQOs apparently probe the Weyl fermions in the *bulk* of TaAs.

To further explore the underlying physics of the giant $\kappa(B)$ MQOs, we now compare $\kappa_{\parallel}(B, B||c)$ and $\kappa_{\perp}(B, B||a)$ measured on the same sample TAc, see Fig. 5. Except for the giant MQOs observed only in $\kappa_{\parallel}(B)$, one recognizes that the smooth background of $\kappa_{\parallel}(B)$ is strongly enhanced by field, as shown by a dashed line (black) denoted as $\kappa_{bg}(B)$. By contrast, $\kappa_{\perp}(B)$ changes only weakly with field, with hardly detected MQOs and a field-induced, weakly decreasing $\kappa_{bg}(B)$. The field-induced decrease of $\kappa_{bg}(B)$ observed in perpendicular field tracks the common field dependence of both $\kappa_{e}(B)$ or $\kappa_{bi}(B)$. Recently, $\kappa_{ph}(B)$ has also been demonstrated to decrease with field [33], due to a field-induced local diamagnetism that exerts an enhanced anharmonic magnetic force on the lattice vibrations. The deceasing ultrasound echo amplitude shown Fig. 4, right axis, indicates that this scenario may apply to TaAs, too. Conversely, the smooth increase of $\kappa_{bg}(B)$ observed for parallel field configuration (Fig. 5) supports the notion of exotic heat carriers that tend to carry an increasing amount of heat with increasing magnetic field.

Below we propose that the CZS of Weyl fermions is a strong candidate to explain the exotic behavior of $\kappa(B)$ that is observed in TaAs [23]. Different to the sound in a solid due to lattice vibrations, CZS in Weyl semimetal is a unique acoustic wave formed by the anti-symmetric combination of the oscillating currents from different Weyl valleys caused by the chiral magnetic effect. As discussed in ref. 23, the CZS is a collective bosonic excitation of Weyl fermions and will propagate along the magnetic field, without any net charge current involved. Its velocity is anti-proportional to the electronic DOS at the Fermi level, leading to strong quantum oscillations for the thermal conductivity along the magnetic field. For a perpendicular field configuration, the CZS is expected to affect the thermal conductivity to some extent, too, due to its contribution to the specific heat [23]. Furthermore, the non-oscillatory part of the thermal conductivity due to CZS along the field will increase linearly with field *B*, as the sound velocity is proportional to *B*. These distinct features of the CZS appear to be well consistent with the main experimental results reported in this work, i.e., the giant and much smaller MQOs in $\kappa_\parallel(B)$ and $\kappa_\perp(B)$, respectively, as well as the increasing background $\kappa_{bg}(B)$ in the former.

In Fig. 5, we also present a semiquantitative fitting to our experimental results based on the CZS scenario. There are three relevant material parameters in the CZS theory [23]: the Fermi wavevector $k_F$, the quasiparticle lifetime $\tau_0$, and the intervalley relaxation time $\tau_v$. The former two parameters can be determined experimentally from the frequency of quantum oscillations and the DC electric conductivity, respectively. By choosing a proper intervalley relaxation time according to the relative amplitude of the quantum oscillations in thermal conductivity, we can calculate $\kappa_\parallel(B)$ by using Eq. 11 in SM [24]. As shown in Fig. 5, the calculated results (green solid line) can reproduce the experimental values reasonably well; see SM for a detailed note on the calculation [24]. Furthermore, we indeed find $\tau_v \gg \tau_0$, which suggests that the intravalley scattering is much stronger than the intervalley scattering and therefore the CZS is truly stabilized by the so-called chiral limit [23,24].

As the strength of the *e-p* interaction is proportional to (or, in other words, the resulted phonon relaxation time is anti-proportional to) the quantized DOS, one expects the MQOs of the common phonon contribution $\kappa_{ph}(B)$ to be inphase with and additive to those of the CZS. This contribution may partly explain the discrepancy of the semiquantitative fitting shown in Fig. 5. Obviously, our ultrasound measurements have detected an acoustic phonon, as the observed phonon velocity $\upsilon_{ph} \approx 2010$ m/s is a typical value for transverse acoustic phonons in TaAs [34]. In our ultrasound spectra we could not directly detect the CZS which is expected to occur at high frequencies (a rough estimate based on the dominating intervalley relaxation time $\tau_v$ indicates a frequency of at least several hundred MHz). Another aspect concerns the sound velocity of CZS, which is estimated to be 5 x $10^4$ m/s for typical Weyl semimetals [23], one order of magnitude larger than that of acoustic phonons in common solids. Therefore, detection of the CZS mode by ultrasound technique requires much higher time resolution; the efforts in this direction are still underway.

In conclusion, we have observed giant magnetic quantum oscillations in the magneto-thermal conductivity of TaAs. All the conventional mechanisms due to diffusion of electrons, phonons and bipolar excitations are unable to account for these phenomena. Alternatively, we have identified that chiral zero sound associated with the Weyl fermions is the most likely source for these observations. This is a collective bosonic excitation of Weyl fermions which acts as an efficient heat carrier and results in a huge ratio of heat over charge conduction in a finite temperature regime around a few Kelvins. It is worth mentioning that, except for the fundamental significance, the large thermal conductance observed in this work may also find feasible applications in heat management of electronic devices as, e.g., a controllable heat switch.

**Methods**

The thermal conductivity was measured utilizing a home-designed sample holder with a chip resistor of 2000 $\Omega$ as heater and a thin ($\phi$ = 25 $\mu$m) chromel-AuFe$_{0.07\%}$ thermocouple for detecting the temperature difference d$T$ across the sample, see Fig. 1a. Silver epoxy H20E (EPO-TEK) was used to make good thermal contact between the sample and the cold plate of the sample holder. The connection points of the thermocouple wires, after being wrapped by a tiny drop of Stycast 2850 (Emerson & Cuming), were pressed onto the sample surface and further fixed by GE varnish in order to ensure thermal connection but electrical insulation to the sample. Thin nylon wires were used to support the thermocouple and heater. Electrical insulation between the sample and the thermocouple had been confirmed before thermal transport measurements. In order to minimize heat loss, thin manganin wires ($\phi$ = 25 $\mu$m) were used as lead wires for heater and thermocouple; and the whole measurement unit was kept in a high vacuum. Furthermore, prior to our measurements, the thermocouple had been carefully calibrated in magnetic fields by using two Cernox CX1050 thermometers; a similar field dependence as previously reported [35] has been observed, cf. Fig. S1 [24].

In order to double-check our observations of the giant $\kappa(B)$ MQOs, we have also employed the thermal transport option equipped on the Physical Properties Measurement System (PPMS, Quantum Design) to measure the magneto-thermal conductivity. This method employs two field-calibrated Cernox thermometers to detect d$T$, see Fig. S2a inset [24]. As shown in Fig. S2a, the thermal conductivity $\kappa_{\parallel}(B)$ obtained by the two different methods agree reasonably well. We note, however, that the thermal measurements by two thermometers are extremely time consuming due to the long relaxation time for each measurement point. By contrast, employment of a thin thermocouple ensures a rapid thermal relaxation during a scan of magnetic field, which is of uttermost importance for reliable detection of the $\kappa(B)$ MQOs. The reliability of the thermal transport measurements can be confirmed from the recorded d$T(B)$ profile at $T$ = 2 K (cf. Fig. 1b), which is almost field-symmetric with reproducible MQOs for both positive and negative fields. Thermal conductivity $\kappa$ is obtained via $\kappa$ = $A$ $Q$/d$T$, where $A$ is the sample geometrical factor and $Q$ the heating power. The measured d$T$ in the whole field range is stable within 5%.

We have also performed ultrasound measurements by using a phase-comparison technique [36] to examine the quantum oscillations of (the velocity and amplitude of) propagating acoustic phonons. $LiNbO_3$ piezoelectric plates with a fundamental resonance frequency of 19 MHz were used as ultrasonic transducers to generate and detect the ultrasound signal, see Fig. 4, inset. The relative change of the sound velocity in the sample can be monitored by the frequency change of the propagating phonons; the corresponding amplitude of the transmitted phonons was directly measured as a voltage generated by ultrasound echoes,

## Acknowledgements

We would like to acknowledge fruitful discussions with G. Li, Y. Li, K. Behnia, H. Weng and Z. Fang, and technical assistance in ultrasound measurements from M. Yoshizawa. This work was supported by the National Science Foundation of China (Grant Nos:11474332, 11474015, 11774018 and 61227902), the MOST of China (Grant Nos: 2015CB921303 and 2017YFA0303103) and the Chinese Academy of Sciences through the strategic priority research program (XDB07020200).

## Figures & Captions

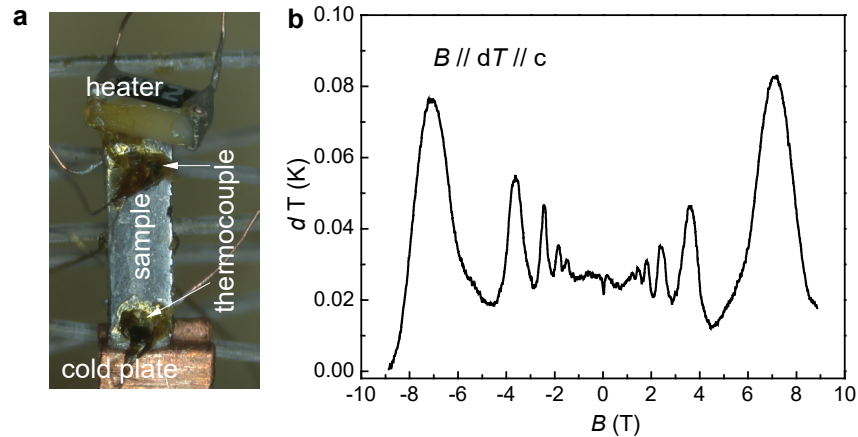

**FIG. 1.** Thermal conductivity measurement configuration and raw experimental data of temperature gradient d$T(B)$. (**a**) TaAs sample (TAc) mounted onto a copper cold plate. A chip resistor of 2000 $\Omega$ was used as heater and a thin ($\phi$ = 25 $\mu$m) chromel-$AuFe_{0.07\%}$ thermocouple for detecting d$T$. Manganin wires ($\phi$ = 25 $\mu$m) used for simultaneous electrical resistivity measurements were connected to the backside of the sample by spot welding. See 'method' for more details on the thermal transport measurements. (**b**) Raw experimental values of d$T$ recorded in a continuous field scan between $B$ = -9 T and 9 T at $T$ = 2 K, with a fixed heating power. Both heat current and magnetic field were applied along the *c* axis.

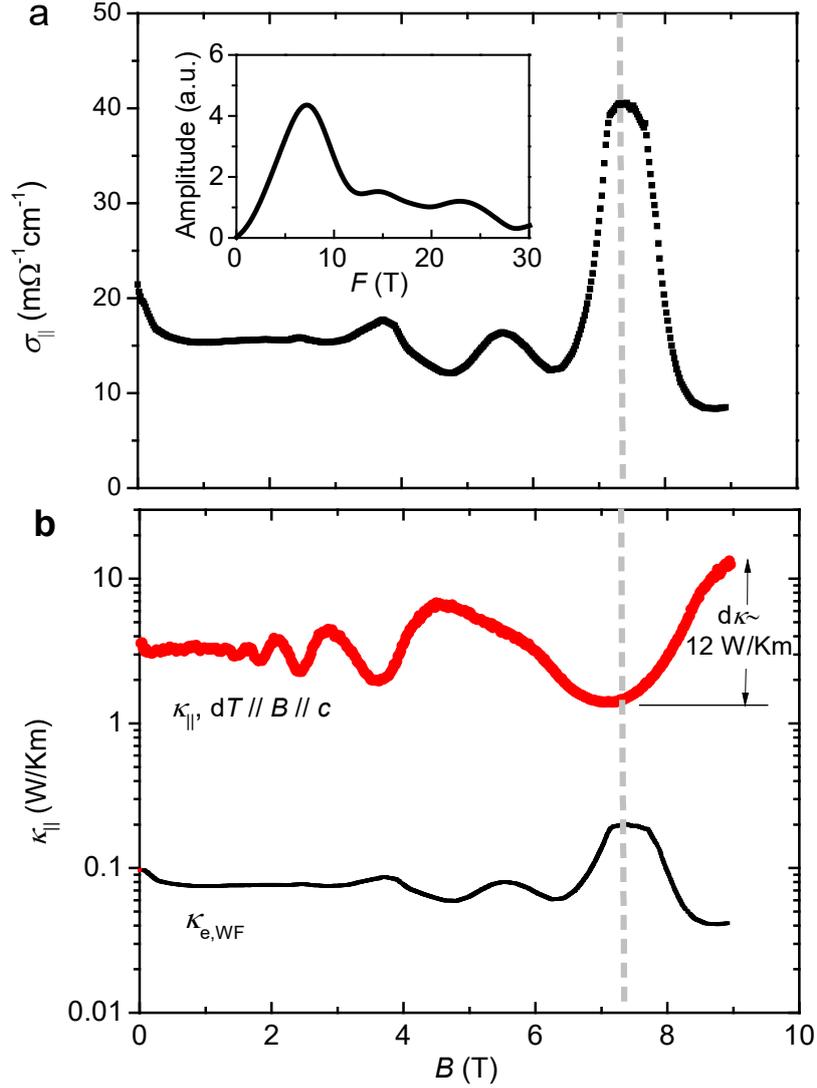

**FIG. 2.** Comparison of quantum oscillations in $\sigma_\parallel(B)$ (**a**) and $\kappa_\parallel(B)$ (**b**) measured at 2 K for sample TAc. In a parallel field configuration, $\sigma_\parallel(B) = 1/\rho_\parallel(B)$ due to the absence of Hall conductivity. The SdH oscillations in $\sigma_\parallel(B)$ reveals three characteristic oscillation frequencies, see FFT analysis shown in the inset, in full agreement with literature data [19,21-22]. The MQOs of $\kappa_\parallel(B)$ (**b**) measured in the parallel field configuration have one dominating frequency of $F \approx 7$ T. Also shown in (**b**) is the calculated $\kappa_{e,WF}(B)$ from $\sigma_\parallel(B)$ using the WF law. The oscillation amplitude of $\kappa_{e,WF}(B)$ is two orders of magnitude smaller than that of the measured ones (red points in **b**). Note the logarithmic vertical axis in panel **b**. The dashed line marks the position of the last extremum, where the measured $\kappa_\parallel(B)$ exhibits a minimum, whereas $\sigma_\parallel(B)$ and the calculated $\kappa_{e,WF}(B)$ assume a maximum, with opposite phases (phase difference $\pi$).

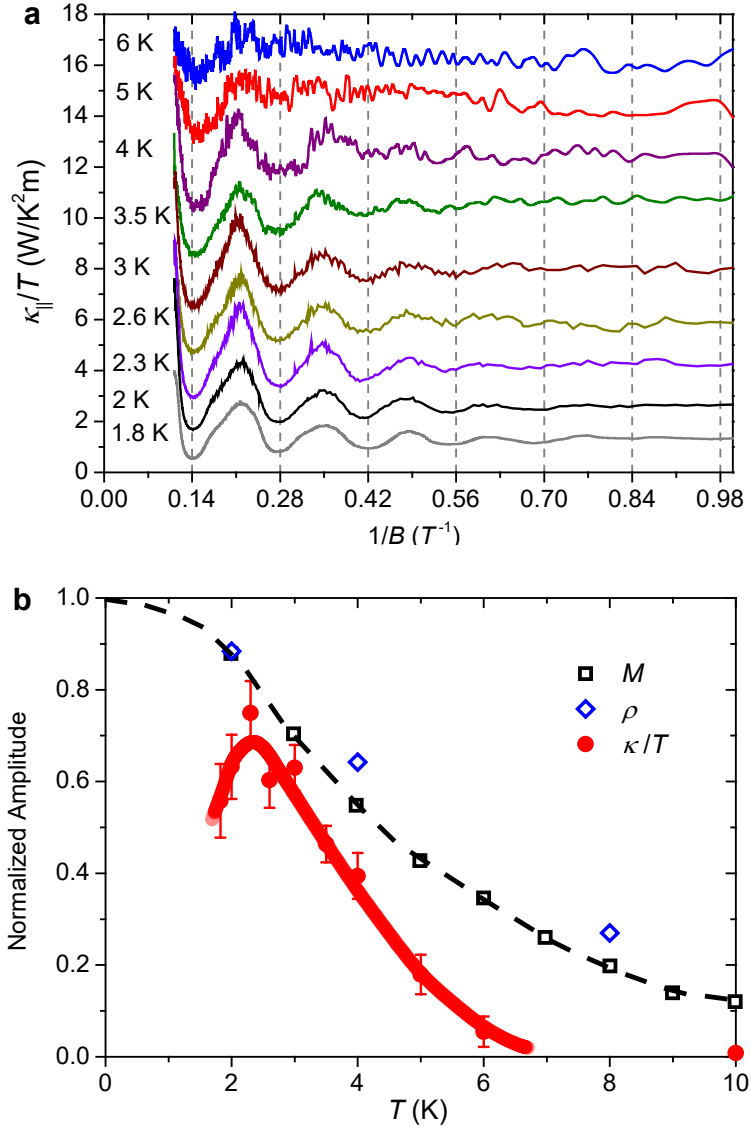

**FIG. 3.** Evolution of the $\kappa_\parallel(B)$ MQOs as a function of temperature. (**a**) Isothermal $\kappa_\parallel/T$ curves as a function of $B^{-1}$ measured at selected temperatures between $T$ = 1.8 and 6 K. A characteristic frequency $F \approx 7$ T can be readily observed from the $T$-independent oscillation period of ~0.14 T$^{-1}$ for all the curves. The corresponding FFT analysis is shown in Fig. S7. For the sake of clarity, the $\kappa_\parallel/T$ curves have been shifted upward by 7 (6 K), 6 (5 K), 5 (4 K), 5.5 (3.5 K), 4.5 (3 K), 3.5 (2.6 K), 2 (2.3 K), 1 (2 K) units of W/K$^2$m, respectively. (**b**) Temperature dependence of the normalized FFT amplitude of the characteristic MQOs with $F \approx 7$ T, compared to the ones observed in magnetization (extracted from ref. 22) and electrical conductivity (cf. Fig. 2a).

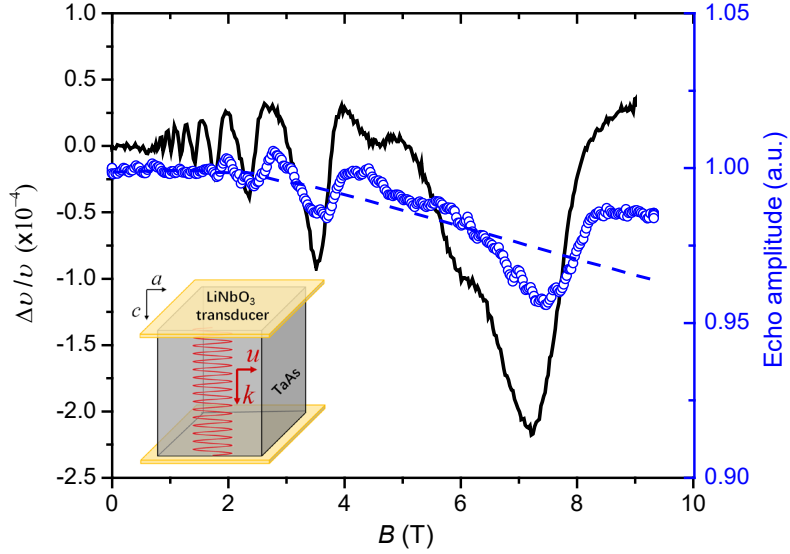

**FIG. 4.** Ultrasound velocity (left) and the corresponding ultrasound echo amplitude (right) measured as a function of field at $T = 2$ K. A transverse sound of 19 MHz was used, with the propagation ($k$) and polarization ($u$) along the $c$ and $a$ axes, respectively (see inset). The detected sound velocity is approximately 2010 m/s, as expected for the transverse acoustic phonons of TaAs [34]. These measurements were performed under the condition that the ultrasound propagation and magnetic field were parallel and along $c$ axis, corresponding to the field configuration of $\kappa_\parallel(B)$ measurements. The dominating oscillations in both quantities have a frequency of 7 T, almost in phase with $\kappa_\parallel(B)$ MQOs, see Fig. 2b. The quantum oscillations of the sound velocity are of the order of $1/10^4$ up to 9 T, whereas the corresponding echo amplitude oscillates within 5% of its zero field value. The dashed blue line indicates the decreasing background of the ultrasound echo amplitude as a function of $B$, which implies a decreasing phonon mean-free path.

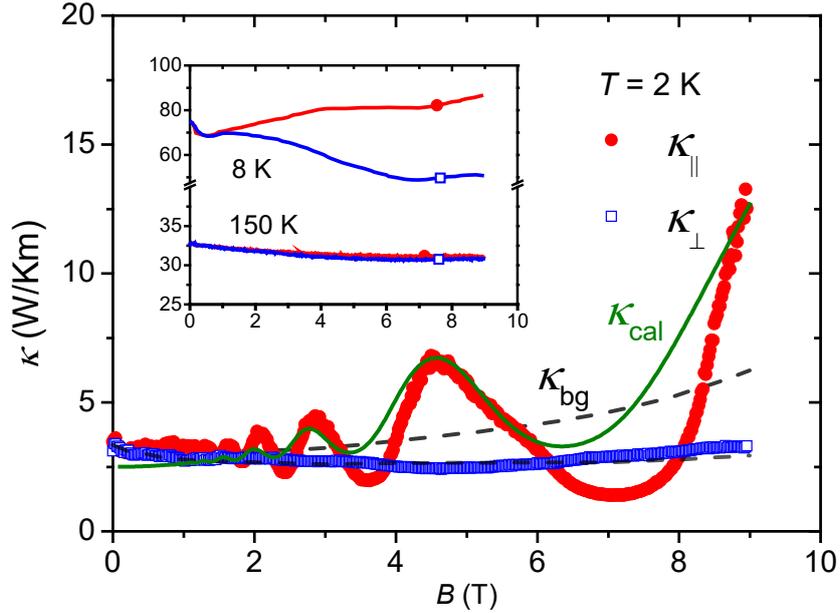

**FIG. 5.** Comparison of $\kappa(B)$ measured for sample TAc in parallel ($dT\|c$, $B\|c$) and perpendicular (d$T$||$c$, $B$||$a$) field configurations. The dashed black lines denote a smooth background $\kappa_{bg}(B)$ for $\kappa_{||}(B)$ and $\kappa_{\perp}(B)$. It is clearly seen that $\kappa_{bg}(B)$ for the parallel field configuration gradually increases, whereas that for the perpendicular field slightly decreases with $B$. These apparently different trends of $\kappa_{bg}(B)$ between the two field configurations provide strong support for CZS being involved in $\kappa_{||}(B)$. Furthermore, a semiquantitative fitting based on the CZS scenario is also shown (solid green $\kappa_{cal}(B)$ line), which agrees reasonably well with the experimental results. Inset: The trend of the background contribution for the two different field configurations shown in the main panel is much better demonstrated at $T = 8$ K, where the $F \approx 7$ T oscillations have almost disappeared, cf. Fig. 3b. At a much higher temperature ($T = 150$ K), the data for $\kappa_{||}(B)$ and $\kappa_{\perp}(B)$ fall on top of each other.